\documentclass[conference]{IEEEtran}
\usepackage{amsmath}

\usepackage[]{graphicx,color}
\usepackage{epstopdf}
\usepackage{amsmath,amssymb}
\usepackage{bm}
\usepackage{url}
\usepackage{hyperref}
\usepackage{ascmac}

\usepackage{algorithm}
\usepackage{algpseudocode}
\makeatletter
\newcommand{\removelatexerror}{\let\@latex@error\@gobble}
\makeatother

\usepackage{eqparbox}
\newdimen{\algindent}
\algnewcommand\LeftComment[2]{%
\hspace{#1\algindent}$\triangleright$ \eqparbox{COMMENT}{#2} \hfill %
}

\algnewcommand\algorithmicinput{\textbf{Input:}}
\algnewcommand\INPUT{\item[\algorithmicinput]}
\algnewcommand\algorithmicoutput{\textbf{Output:}}
\algnewcommand\OUTPUT{\item[\algorithmicoutput]}

\IEEEsettopmargin{t}{0.8in}

\usepackage{comment}

\title{{Trainable Projected Gradient} Detector\\ for Sparsely Spread Code Division Multiple Access}
\author{
  \IEEEauthorblockN{Satoshi Takabe\IEEEauthorrefmark{1}\IEEEauthorrefmark{2}, 
 		Yuki Yamauchi\IEEEauthorrefmark{1}, 
                Tadashi Wadayama\IEEEauthorrefmark{1}}
  \IEEEauthorblockA{\IEEEauthorrefmark{1}%
		Nagoya Institute of Technology,
		Gokiso, Nagoya, Aichi 466-8555, Japan,\\
 		s\_takabe@nitech.ac.jp,  y.yamauchi.475@stn.nitech.ac.jp, wadayama@nitech.ac.jp} 
  \IEEEauthorblockA{\IEEEauthorrefmark{2}%
  		RIKEN Center for Advanced Intelligence Project,
  		Nihonbashi, Chuo-ku, Tokyo 103-0027, Japan
                }
}

\begin{document}
%
\maketitle

\begin{abstract}
Sparsely spread code division multiple access (SCDMA) is a promising non-orthogonal multiple access technique for future wireless communications.
In this paper, we propose a novel trainable multiuser detector called sparse trainable projected gradient (STPG) detector, which is based on the notion of deep unfolding.
In the STPG detector, trainable parameters are embedded to a projected gradient descent algorithm, which can be trained by standard deep learning techniques such as back propagation and stochastic gradient descent.
Advantages of the detector are its low computational cost and small number of trainable parameters, which enables us to treat massive SCDMA systems. 
In particular, its computational cost is smaller than a conventional belief propagation (BP) detector while the STPG detector exhibits nearly same detection performance with a BP detector.
We also propose a scalable joint learning of signature sequences and the STPG detector for signature design. 
Numerical results show that the joint learning improves multiuser detection performance particular {in the low SNR regime}.

\end{abstract}


\section{Introduction}\label{sec_intro}

Non-orthogonal multiple access (NOMA) is a key ingredient of recent multiple access techniques for the fifth generation (5G) mobile networks.
By allocating several users to the same resource block, NOMA techniques realize high spectral efficiency and low latency even when a network is massively  connected~\cite{NOMA}.
Additionally, in future multiple access communications, \emph{overloaded} access is considered to be unavoidable because of spectral resource limitation.
NOMA techniques are expected to deal with such an overloaded system in which the number of active transmitters is larger than that of the signal dimension~\cite{SCMA}.

Code division multiple access (CDMA)~\cite{Hara} is an OMA system in which {$n$ active users} communicate with a base station (BS) simultaneously by spreading users' signals with their  \emph{signature sequences} (or spreading sequences) {of length $m$}.
Although orthogonality of signature sequences ensures a reasonable multiuser detection performance
 {if $m\ge n$, 
detection performance will drop in overloaded cases ($m<n$).}       
\emph{Sparsely spread CDMA} (SCDMA)~\cite{YT} is a promising NOMA technique based on CDMA.
In SCDMA, data streams are {modulated} by randomly generated signature sequences which contain a small number of non-zero elements. 
The BS receives superimposed signals with additive noise and tries to detect data streams from multiple users. 
Compared with conventional CDMA, sparse signature sequences in SCDMA allow low-complexity detection using a linear-time algorithm such as the belief propagation (BP).
Moreover, as a NOMA system,  SCDMA potentially achieves reasonable detection performance even in overloaded cases.

Recent studies on SCDMA mainly focused on design of detectors and signature sequences.
As described above, BP is a detector suitable for the sparse structure of SCDMA~\cite{Guo}, which exhibits nearly optimal performance predicted theoretically~\cite{YT, Tse}. 
{The computational complexity of the BP detector rapidly increases with respect to signature sparsity and 
the constellation size of transmit signals.
Since practical SCDMA systems use sufficiently large values of these parameters, we need to reduce a computational cost for faster multiuser detection. }
Signature design is another crucial issue because detection performance depends on superimposed signals {spreaded by signature sequences}. 
In~\cite{Song}, a signature matrix family that improves BP detection performance is proposed.
Recently,~\cite{Kim} and~\cite{Lin} proposed an alternative approach for related SCMA systems which designs signature sequences and a detector jointly by autoencoders.
Although learned autoencoders provide signature sequences with reasonable performance, 
their high training cost is a drawback because they contain a large number of training parameters. 
In summary, a desirable detector and signature design should posses both high scalability for large systems and good adaptability to practical SCDMA systems with high signature sparsity, large signal constellations, and/or overloaded access.

Rapid development of deep learning (DL) techniques has stimulated design of wireless communication systems~\cite{ML}.
Recently, \emph{deep unfolding} proposed by Gregor and LeCun~\cite{LISTA} has attracted great interests as another DL-based approach~\cite{DU} in addition to an end-to-end approach~\cite{E2E}. 
In deep unfolding, the signal-flow graph of an existing iterative algorithm is expanded to a deep network architecture
 in which some parameters such as step-size parameters are embedded. 
These embedded parameters are treated as trainable parameters to tune the behavior of the algorithm. 
Learning trainable parameters is accomplished by standard supervised learning techniques such as back propagation and stochastic gradient descent (SGD) if the original algorithm consists of differentiable processes. 
An advantage of deep unfolding is that the number of trainable parameters are much fewer than conventional deep neural networks, which leads to fast and stable training process and high scalability.
Deep unfolding has been applied to various topics in signal processing and wireless communications: sparse signal recovery~\cite{LISTA,TISTA,TISTA2}, massive MIMO detection~\cite{He, TPG,TPG2}, 
signal detection for clipped OFDM systems~\cite{CTISTA} and trainable decoder for LDPC codes~\cite{LDPC}.

In this paper, we propose a trainable multiuser detector and signature design with high scalability and adaptability to overloaded  SCDMA systems. 
In order to resolve a scalability issue of multiuser detection, we first introduce a novel SCDMA {multiuser} detector called sparse trainable gradient projection (STPG) detector. 
The STPG detector is based on a projected gradient descent algorithm whose gradient can be computed efficiently. 
Combined with the deep unfolding technique, we will propose a trainable detector with reasonable detection performance, high scalability, and adaptability to practical SCDMA systems.
In addition, a scalable DL-based SCDMA signature design is proposed by learning a signature matrix 
 and STPG detector simultaneously. 
In the proposed method, 
 values of non-zero elements in a signature matrix and trainable parameters of the detector are  jointly trained to improve detection performance based on an estimate from a temporal signature matrix and detector.
Compared with existing DL-based approaches, the proposed method can be trained in huge systems.

The outline of the paper is as follows. 
Section~\ref{sec_2} describes a system model and conventional BP detector.
In Section~\ref{sec_3}, we propose the STPG detector for SCDMA multiuser detection and compare its detection performance in large systems with a BP decoder.  
Section~\ref{sec_4} describes signature design based on STPG detector and demonstrate performance improvement.
Section~\ref{sec_5} is a summary of this paper.

\section{System model and BP detector}\label{sec_2}
We first introduce SCDMA system model and a conventional BP detector.

\subsection{SCDMA system model}

We consider an uplink SCDMA system {where $n$ active users with a single antenna
try to transmit their messages to a BS by using signature sequences of  length $m$}.
The ratio $\beta := n/m$ is called overloaded factor.
From the definition, $\beta>1$ indicates that the system is overloaded, i.e., $m<n$.  
We assume that the ratio $\beta$ is a constant number. 
Each user has a BPSK-modulated signal $x_i\in\{+1,-1\}$ ($i=1,\dots,n$) as a transmit data.
In addition, users have their own signature sequences $\bm{a}_i=(a_{1,i},\dots,a_{m,i})^T\in\mathbb{R}^m$.
Then, the BS receives superimposed signals given by
\begin{equation}
\bm{y} = \sum_{i=1}^n \bm a_{i}x_i + \bm{w}, \label{eq_ch1}
\end{equation}
where $\bm w$ is a noise vector and $\bm y\in\mathbb{R}^m$ is a received signal {at the BS}. 
Letting $\bm A:=(\bm a_1, \dots, \bm a_n)\in \mathbb{R}^{m\times n}$ be a signature matrix, (\ref{eq_ch1}) has another form written by
\begin{equation}
\bm{y} = \bm A \bm x + \bm{w}, \label{eq_ch2}
\end{equation}
where $\bm x := (x_1,\dots, x_n)^T$.
In {conventional} CDMA systems, we assume orthogonality of signature sequences $\bm{a}_i^T\bm{a}_j=0$ for any $i\neq j$. 
Instead, SCDMA systems require sparsity of signature sequences so that the number of non-zero elements in each signature sequence is constant to $n$ and $m$. 

We consider the following typical SCDMA system.
First, we assume an AWGN channel. 
Second, each row of the signature matrix $\bm A$ is assumed to have $k$ non-zero entries, which is called signature sparsity in this paper.
We also assume that the signature matrix is normalized such as $\|\bm A\|_F^2=km$,
where $\|\cdot\|_F$ {denotes} a Frobenius norm. 
Under these assumptions, the signal-to-noise ratio (SNR) of the system defined by $n_0:=\mathsf{E}_{\bm x}\|\bm A\bm x\|_2^2/\mathsf{E}_{\bm w}\|\bm w\|_2^2$ is
calculated as $n_0=k/\sigma^2$,
where $\sigma^2$ is the variance of the noise per a symbol.
Equivalently, the SCDMA model for a given SNR $n_0$ is defined by
\begin{equation}
\bm{y} = \sqrt{\frac{n_0}{k}}\bm A \bm x + \bm{w}_0, \label{eq_ch3}
\end{equation}
where $\bm{w}_0$ is an i.i.d. Gaussian random vector with zero mean and unit variance.
We consider a multiuser detector and signature design for this system model.

\subsection{BP detector}\label{sec_BP}
We briefly describe a BP detector of a standard multiuser detector for SCDMA~\cite{Guo}.

Recursive equations of the BP are constructed on a {factor} graph whose nodes are variables $\bm x$ and $\bm y$ and edges are set according to non-zero elements of $\bm{A}$.
The message $U_{j\rightarrow i}(x)$  ($x\in\{+1,-1\}$) is a message from a {chip} node $y_j$ to a {symbol} node $x_i$, and $V_{i\rightarrow j}(x)$ is a message from a {symbol} node $x_i$ to a {chip} node $y_j$.
Then, the BP recursive formula is given by
\begin{align}
V_{i\rightarrow j}(x) &= Z_{i\rightarrow j}^{-1}\prod_{l\in\partial i\backslash j} U_{l\rightarrow i}(x),\\
U_{j\rightarrow i}(x) &= Z_{j\rightarrow i}^{-1}\sum_{\bm{x}_{\partial j\backslash i}}
\left(\prod_{k\in \partial j\backslash i} V_{k\rightarrow j}(x_k)\right)\nonumber\\
\times& \exp\left\{-\frac{1}{2}\left[y_j-\sqrt{\frac{n_0}{k}}\left(a_{j,i}x+\sum_{k\in \partial j\backslash i}a_{j,k}x_k\right)\right]^2 \right\}, \label{eq_BP}
\end{align}
where $Z_{i\rightarrow j}$ and $Z_{j\rightarrow i}$ are normalization constants, and  
$\partial i:=\{j\in \{1,\dots,m\}| a_{j,i}=1\}$ and $\partial j:=\{i\in \{1,\dots,n\}| a_{j,i}=1\}$ are neighboring node sets on the {factor} graph.
After $T_{\mathrm{BP}}$ iterations, the probability that the $i$th transmit signal takes $x$ is estimated by  
\begin{equation}
V_{i}(x) = Z_{i}^{-1}\prod_{j\in\partial i} U_{j\rightarrow i}(x),
\end{equation}
where $Z_{i}$ is a normalization constant. 
Finally, the $i$th transmit signal is detected as 
$x_i=1$ if  $V_{i}(1)\ge V_{i}(-1)$, and $x_i=-1$ otherwise.

{The computational cost of the BP detector is $O(k^22^{k-1}n)$ because (\ref{eq_BP}) contains a sum over all possible combinations of $k-1$ transmit signals $\bm{x}_{\partial j\backslash i}$. 
Similarly, if the detector is applied to a system with a higher order modulation of size $|\mathcal M|$,
 the computational cost of the BP detector is $O(k^2|\mathcal{M}|^{k-1}n)$.
This rapid increase with respect to $k$ and $|\mathcal{M}|$ is a drawback of the BP detector. 
}

\section{STPG detector}\label{sec_3}
In this section, we propose a trainable multiuser detector for SCDMA using the idea of deep unfolding.

\begin{table*}[t]
  \begin{center}
    \caption{Number of operations in STPG and BP detectors, and values for various $k$ when  $n=m=1200$ ($\beta=1$).}
    \begin{tabular}{|c|c|c|c|c|c|} \hline
       & Number of operations & $k=2$ & $k=4$ & $k=6$ & Big-O notation \\ \hline \hline
      STPG additions & $(2\beta^{-1}k + \beta^{-1}+1)n$ & $7.20\times 10^3$ &  $1.20\times 10^4$
      & $1.68\times 10^4$ & $O(kn)$ \\
      BP additions& $(k2^k + 2)\beta^{-1}kn$ & $2.40\times 10^4$ & $3.16\times 10^5$ 
      & $2.77\times 10^6$ & $O(k^22^kn)$ \\ \hline
      STPG multiplications & $( \beta^{-1}k+ \beta^{-1} +2)n+1$ & $6.00\times 10^3$ & $8.40\times 
10^3$ 
      & $1.08\times 10^4$  & $O(kn)$ \\
      BP multiplications & $\{(2 k + 3)2^k + 2 \beta^{-1}k\}\beta^{-1}kn$ & $7.68\times 10^4$ & $8.83\times 10^5$
      & $6.99\times 10^6$  & $O(k^22^kn)$ \\ \hline
    \end{tabular}
    \label{tab_1}
  \end{center}
\end{table*}

\subsection{Structure of STPG detector}

Deep unfolding is an efficient DL-based approach borrowing the structure of an  iterative algorithm.
Here, we employ a gradient descent-based detector apart from a message-passing algorithm such as a BP.

The maximum-likelihood (ML) estimator for SCDMA system (\ref{eq_ch3}) is formulated by
\begin{equation}
\bm{\hat x} = \mathrm{argmin}_{\bm{x}\in\{+1,-1\}^n} \left\|\bm{y}-\sqrt{\frac{n_0}{k}}\bm A\bm x\right\|_2^2. \label{eq_ml}
\end{equation}
This ML estimator is formally similar to that for the MIMO signal detection~\cite{TPG},
and computationally intractable for large $n$.
Alternatively, a projected gradient descent  (PG) algorithm is used to solve (\ref{eq_ml}) approximately by replacing the constraint $\bm{x}\in\{+1,-1\}^n$ to relaxed one $\bm{x}\in[-1,1]^n$. 
Its recursive formula of the PG is given by
\begin{align}
\bm r_t &= \bm s_t + \gamma \sqrt{\frac{n_0}{k}}\bm A^T\left(\bm y- \sqrt{\frac{n_0}{k}}\bm A \bm s_t\right), \label{eq_pg_1}\\
\bm s_{t+1} &= \tanh(\alpha \bm r_t), \label{eq_pg_2}
\end{align}  
where $\bm{s}_0=\bm 0$ is an initial vector.
The first equation is called a gradient step because $\bm r_t$ is updated by a gradient descent method with a step size $\gamma>0$.
The next equation is named a projection step with an element-wise soft projection function $\tanh(\cdot)$.
The softness parameter $\alpha$ controls the shape of the soft projection function. In the large-$\alpha$ limit, the function becomes a step function, which is the original projection function onto $[+1,-1]$. 
It is expected that the detection performance of the PG depends on the choice of the parameters $\gamma$ and $\alpha$.
{As a disadvantage of the plain PG, we should search values of parameters carefully for reasonable performance.}

\begin{figure}[t]
  \centering
  \includegraphics[width=7.5cm]{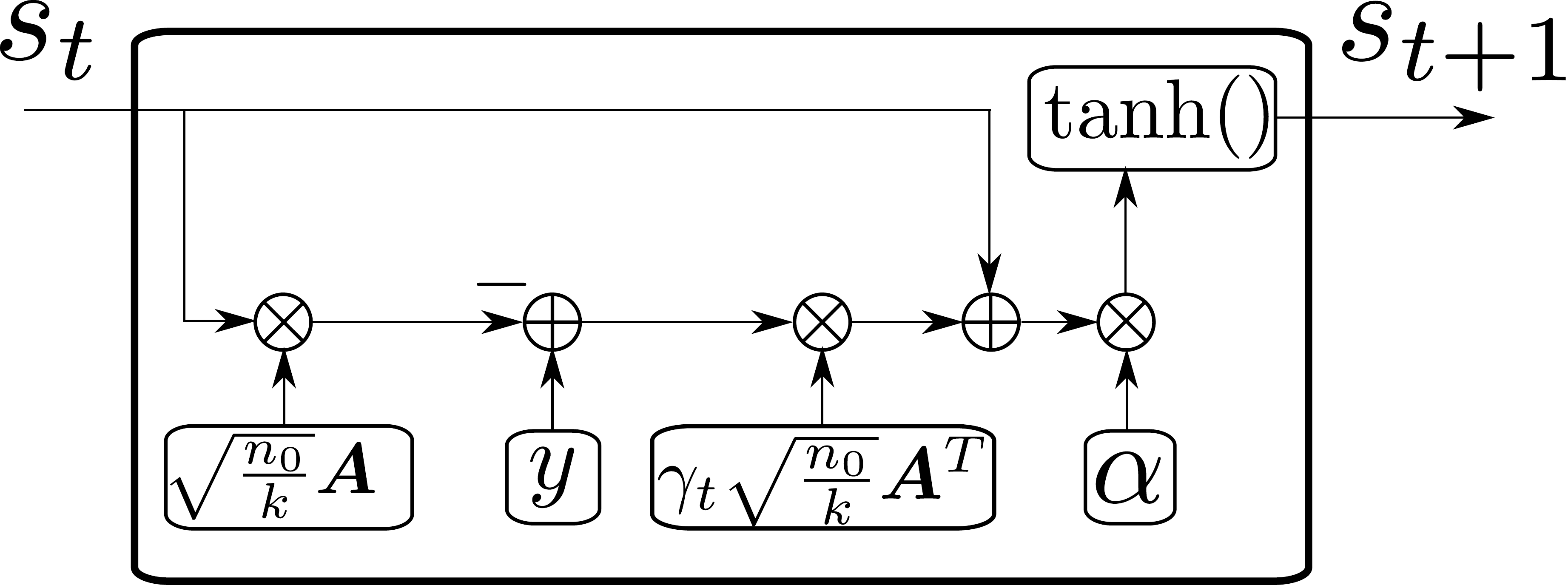}
  \caption{Architecture of the STPG detector at the $t$th iteration.}
  \label{fig_ar}
\end{figure}

To introduce the STPG detector, we replace a parameter $\gamma$
to $\gamma_t$~\footnote{The parameter is introduced as {$\gamma_t^2$} to avoid a negative step size in implementation. } depending on the iteration step $t$.
The proposed STPG detector is thus defined by
\begin{align}
\bm r_t &= \bm s_t + \gamma_t \sqrt{\frac{n_0}{k}}\bm A^T\left(\bm y- \sqrt{\frac{n_0}{k}}\bm A \bm s_t\right), \label{eq_tpg_1}\\
\bm s_{t+1} &= \tanh(\alpha \bm r_t), \label{eq_tpg_2}
\end{align}  
where $\{\gamma_t\}_{t=1}^T$ and $\alpha$ are regarded as trainable parameters.
The architecture of the $i$th iteration of the STPG detector is shown in Fig.~\ref{fig_ar}.
Note that, although the trainable parameter $\alpha$ can be replaced to $\alpha_t$, 
a single trainable parameter $\alpha$ is used here to reduce the number of trainable parameters.
The total number of trainable parameters is $T+1$ in $T$ iterations, which is constant to $n$ and $m$. This leads to high scalability and stable convergence in its training process.

It is also emphasized that the STPG detector uses $\bm A^T$ in the gradient step although
a similar MIMO detector called TPG-detector uses the pseudo-inverse matrix $\bm U:=( \bm A^T\bm A)^{-1}\bm A^T$~\cite{TPG} or $\bm U_\eta:=(\bm I+\eta \bm A^T\bm A)^{-1}\bm A^T$ with a trainable parameter $\eta$~\cite{TPG2}.
This change reduces the computational complexity of the detector.
In particular, the sparse structure of a signature matrix $\bm A$ in SCDMA enables us to calculate all the matrix-vector product operations in $O(n)$ time. 
On the other hand, even though $\bm A$ is a sparse matrix, a matrix-vector product including $\bm U$ or $\bm U_{\beta}$ takes $O(n^2)$ operations because these matrices are dense in general.  
The details of the computational cost is described in the next subsection.

\subsection{Computational complexity}\label{sec_com}

A crucial property of SCDMA is a low computational cost in multiuser detection.
{Here, we count the number of additions and multiplications of the STPG  and BP detectors in each iteration step.}
For simplicity, we neglect a calculation of nonlinear functions such as $\tanh(\cdot)$ in the STPG detector and $\exp(\cdot)$ in the BP detector.

Table~\ref{tab_1} shows the number of operations {per an iteration} as a function of $n$, $\beta$ and $k$.
In addition, we show the values when $n=m=1200$ ($\beta=1$) and {$k=2,4,6$} for comparison.
It is found that both detectors are linear-time algorithms with respect to $n$.
In particular, the use of $\bm A^T$ in a gradient step helps the STPG detector to reduce its complexity.

We also find that the STPG detector requires less number of operations than the BP detector in terms of {signature sparsity $k$.}
In fact, the STPG detector has $O(kn)$ additions/multiplications in each iteration. On the other hand, the BP detector needs $O(k^22^kn)$ operations {as discussed in Sec.~\ref{sec_BP}.}
As shown in Fig.~\ref{fig_3}, the constant $k$ should be large enough to ensure reasonable detection performance, which results in the rapid increase of the BP computational cost.

It is noteworthy that the gap of computational complexity in terms of $k$ will increase {if we consider higher order modulations}.
For a constellation of size $|\mathcal M|$, the number of operations of the BP detector is $O(k^2|\mathcal{M}|^{k-1}n)$ while that of the STPG detector remains $O(kn)$.
This is a strong point of the STPG detector for practical SCDMA systems.

\subsection{Simulation settings}\label{sec_ex}

In the following subsections, we compare the proposed STPG detector to the original PG and the BP detector in terms of multiuser detection performance.

In the numerical simulations, we consider a massive SCDMA system with $n=1200$ active users.
A signature matrix $\bm A$ is randomly generated by an element-wise product $\bm A= \bm H\odot \bm W$ where $\bm H\in \{0,1\}^{m\times n}$ is a \emph{mask matrix} and 
 $\bm W\in \mathbb{R}^{m\times n}$ is a weight matrix.
In numerical simulations, each weight of $\bm W$ is uniformly chosen from $\{+1,-1\}$.
The mask matrix $\bm H$ is also randomly generated by Gallager's construction~\cite{Gal} 
{so that its row and column weights are exactly equal to $k$ and $k'=km/n(\in\mathbb{N})$, respectively.}

For the PG and STPG detectors, we set $T=30$ as a number of iterations.
The STPG detector is implemented by PyTorch 1.2~\cite{PyTorch}.
{Initial values of trainable parameters are set to $\gamma_t=0.01$ ($t=1,\dots,T$) and $\alpha=2$.}
In training process of the STPG detector, we can use a mini-batch training by back propagation and SGD. 
In addition, the use of incremental training~\cite{TISTA2, TPG2} is crucial to avoid a vanishing-gradient problem and obtain reasonable results.
In the incremental training, we begin with learning the trainable parameters $\gamma_1,\alpha$ assuming that $T=1$. This is called the first generation of training.
After the first generation is finished, we next train parameters $\gamma_1,\gamma_2,\alpha$ as if $T=2$ by using the trained values of $\gamma_1,\alpha$ as their initial values.
Learning these generations is repeated in an incremental manner until $T$ reaches to the desired value.  
In the following simulations, we use $100$ mini-batches of size $200$.
We use the Adam optimizer~\cite{Adam} whose learning rate is $0.0005$.
Training process of the detector is executed for each SNR.
 
Multiuser detection performance is measured by bit error rate (BER).
Since outputs $\bm{s}_{T}$ of the PG and STPG detectors are continuous values, 
a sign function $\mathrm{sign}(x)=1$ ($x\ge 0$) and $-1$ ($x<0$) is applied to the outputs. Thus, the detected signal is given by $\bm{\hat x}=\mathrm{sign}(\bm s_T)$.

\subsection{Acceleration of convergence in STPG}

\begin{figure}[t]
  \centering
  \includegraphics[width=8cm]{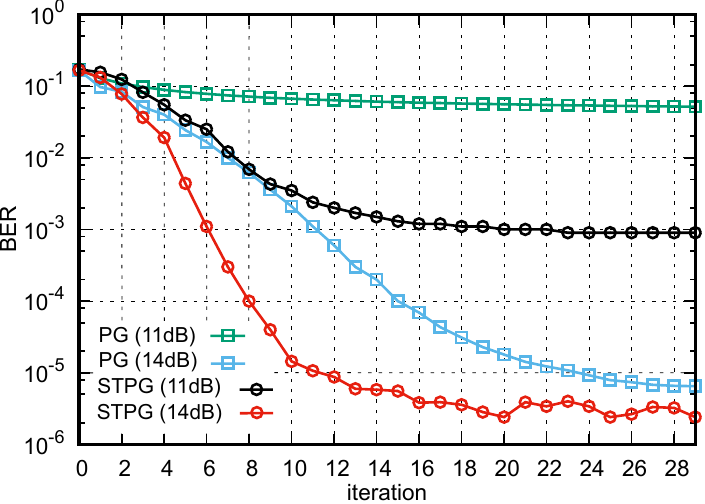}
  \caption{{BER of the PG and STPG detectors as a function of the number of iterations $T$ with different SNRs; $n=1000$, $m=1200$ ($\beta=1.2$), and $k=6$. Parameters of the PG are set to $\gamma=0.01$ and $\alpha=2$.}}
  \label{fig_2}
\end{figure}

We first compare the STPG detector to the original PG to demonstrate advantages of learning parameters by deep unfolding.
Figure~\ref{fig_2} shows the BER performance of both detectors with different SNRs.
{In the original PG, we choose $\gamma=0.01$ and $\alpha=2$ corresponding to initial values of the STPG detector}.
We find that the STPG detector exhibits better performance than the PG.
For example, when SNR is 11dB, the BER of the STPG detector ($T=30$) is about $1.0\times 10^{-3}$ while that of the PG is about $5.1\times 10^{-2}$. 
In addition, when {SNR}$=$14dB, the STPG detector shows fast convergence to a fixed point compared with the PG.
These results indicate that training a constant number of parameters in the PG leads to 
better detection performance and fast convergence to a fixed point. 
Detection performance improvement and convergence acceleration are crucial advantages of deep unfolding as shown in other signal detection problems~\cite{TISTA2, TPG2}.

\subsection{Performance comparison to BP detector}

\begin{figure}[t]
  \centering
  \includegraphics[width=8cm]{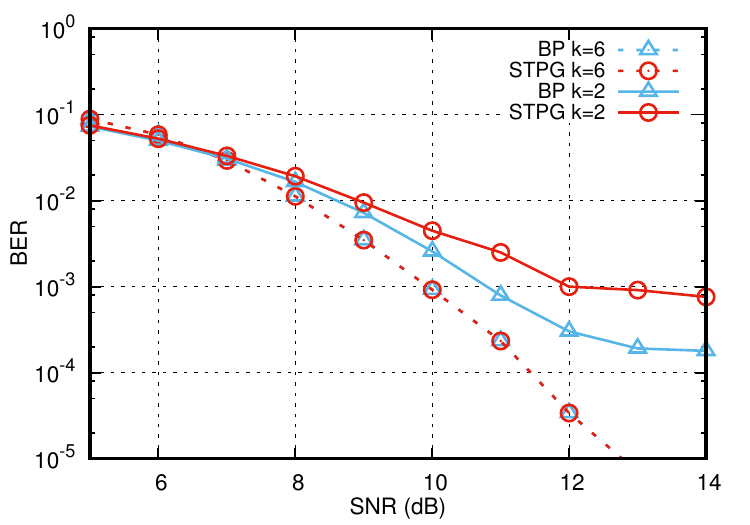}
  \caption{{BER performance of the STPG (circles) and BP detectors (triangles) for SCDMA with signature sparsity $k=2$ (solid line) and $6$ (dotted line); $n=m=1200$ ($\beta=1$).}}
  \label{fig_3}
\end{figure}

\begin{figure}[t]
  \centering
  \includegraphics[width=8cm]{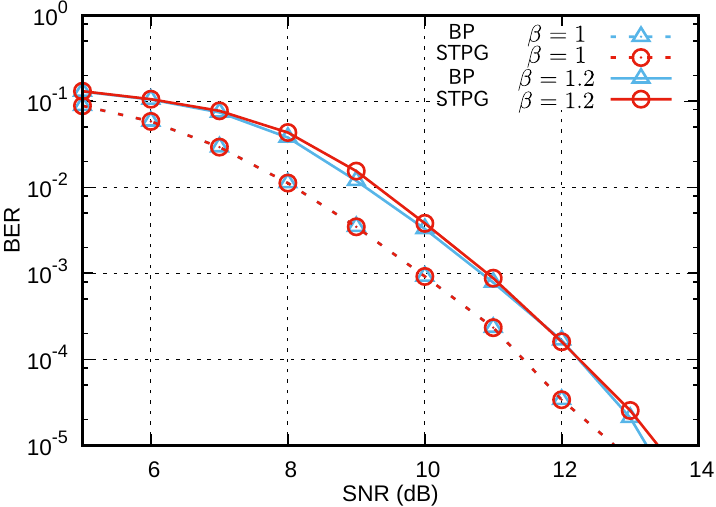}
  \caption{{BER performance of the STPG (circles) and BP detectors (triangles) for SCDMA with overloaded factor $\beta=1$ (dotted line) and $1.2$ (solid line); $n=1200$ and $k=6$.}}  \label{fig_5}
\end{figure}

Next, we compare the STPG detector to a conventional BP detector.

Figure~\ref{fig_3} shows multiuser detection performance of the {STPG ($T=30$) and BP ($T_{\mathrm{BP}}=30$)} detectors {with $n=1200$ active users and $m=1200$ signature sequence length}.
Since $n=m$ ($\beta=1$), reasonable detection performance is expected by using proper signature sequences.
In fact, two detectors exhibit nearly same performance when $k=6$.
When $k=2$, however, the overall BER performance of both detectors decreases while the STPG detector is inferior to the BP detector.
It suggests that the sufficiently large $k$ is preferable for reliable multiuser detection, which leads to a rapid increase of the computational cost of the BP decoder.
{When $k=6$, the computational cost of the BP detector is more than a hundred times as high as that of the STPG detector as shown in Tab.~\ref{tab_1}.} 

{Figure~\ref{fig_5} shows the multiuser detection performance with different overloaded factors $\beta$ when $n\!=\!1200$ and $k\!=\!6$.}
In the overloaded case where $\beta\!=\!1.2$ ($m\!=\!1000$), two detectors exhibit similar BER performance. 
Although the overloaded system suffers from about $1$dB performance degradation compared with the $\beta=1$ case, both algorithms successfully detect transmit signals in the high SNR regime, which is an advantage of SCDMA as NOMA. 
{In overloaded systems, a computational cost of a detector is still crucial because
the signature sparsity $k$ should be sufficiently large as well as the $\beta=1$ case.}

In summary, the STPG detector shows similar detection performance to the BP detector even in an overloaded case.
From the discussion in Sec.~\ref{sec_com}, we can conclude that the STPG detector has an advantage in the computational cost  for  sufficiently large signature sparsity $k$.

\section{Signature design with STPG detector}\label{sec_4}
As described in Sec.~\ref{sec_3}, a trained STPG detector shows reasonable SCDMA multiuser detection performance with low computational complexity. 
Moreover, we can train a signature matrix $\bm A$ combined with the STPG detector. 
In this section, we propose a new signature design by learning a signature matrix and the STPG detector simultaneously.

\subsection{Joint learning of signature matrix and STPG detector}
The structure of deep unfolding enables us to train weights of a signature matrix $\bm A$ by back propagation and SGD.
We show signature design with the STPG detector in Alg.~1.
For simplicity, we train a weight matrix $\bm W$ of a signature matrix with a mask matrix $\bm{H}$ fixed. 
The trainable parameters are then a signature matrix $\bm A= \bm H\odot \bm W$ in addition to  
 $\{\gamma_t\}_{t=1}^T$ and $\alpha$ of the STPG detector.
An update rule for these parameters consists of three steps: (i) calculating a temporal $\bm A$, (ii) generating training data, and (iii) updating trainable parameters.
For step (i), a signature matrix is modified to satisfy the sparsity and normalization conditions that might be broken by the previous parameter update.
In line $4$ of Alg.~1, {signature sparsity $k$ of $\bm{A}$} is recovered by multiplying the masking  matrix $\bm H$ to $\bm A$ updated in the last training step.
The normalization condition $\|\bm A\|_{F}=\sqrt{km}$ is  satisfied after line $5$.  
As step (ii), a mini-batch for a parameter update is generated according to the system model~(\ref{eq_ch3}) with a temporal $\bm A$.
Then, as step (iii), trainable parameters including $\bm A$ are updated to reduce the loss value
 calculated by the mini-batch and the STPG detector with $t$ iterations.
This training step belongs to the $t$th generation of incremental training.

Due to sparse signature sequences and architecture of the STPG detector,
the substantial number of training parameters are $km+T+1$ in total.
It realizes sufficiently fast joint learning. 
In fact, the training process is executed within 20 minutes by a PC with GPU NVIDIA GeForce RTX 2080 Ti and Intel Core i9-9900KF CPU (3.6 GHz).

\begin{figure}[!t]\label{alg}
 \removelatexerror
\begin{algorithm}[H]
 \caption{Joint learning of signature matrix and STPG}
\begin{algorithmic}[1]
 \INPUT $m,n,T,k,n_0$, mini-batch size $bs$, number of mini-batches $B$, 
 mask matrix $\bm{H}$
 \OUTPUT Trained params. $\{\gamma_t\}_{t=1}^T,\alpha, \bm{A}$
 \State Initialize $\{\gamma_t\}_{t=1}^T$, $\alpha$, and $\bm{A}$.
 \For{$t=1$ to $T$}
 \Comment{Incremental training}
   \For{$b=1$ to $B$ }
    \Statex   \LeftComment{1}{(i) Masking and normalization of $\bm A$.}
    \State{$\bm A := \bm A \odot \bm H$}
    \State{$\bm A := (\sqrt{km}/\|\bm A\|_F) \bm A$}
    \Statex   \LeftComment{1}{(ii) Generating training data.}
    \State{Generate $\bm x\in\{\pm1\}^{n\times bs}$ and $\bm{w}$ randomly.}
    \State{Generate $\bm y$ by $\bm y = \sqrt{n_0/k} \bm A \bm x +\bm w$.}
    \Statex   \LeftComment{1}{(iii) Update of training params.}
    \State{Estimate $\bm{\hat x}:= \bm{s}_{t}$ by a temporal STPG detector.}
    \State{Calculate MSE loss between $\bm{x}$ and $\bm{\hat x}$.}
    \State{Update $\{\gamma_t\}_{t=1}^T$, $\alpha$, and $\bm{A}$ by an optimizer.}
   \EndFor
  \EndFor
\end{algorithmic}
\end{algorithm}
\end{figure}

\subsection{Multiuser detection performance}

Now we evaluate the multiuser detection performance of the STPG detectors with/without learning a signature matrix.

In the training process of joint learning, we change the number of mini-batches to $1000$ because the number of trainable parameters increases.
A mask matrix with signature sparsity $k=6$ is generated according to Gallager's method.
Initial values of weights of $\bm A$ are set to one and signature matrices are trained independently for a given SNR.
Other conditions are based on the descriptions in Sec. \ref{sec_ex}.
For the STPG detector with fixed $\bm A$, weights of $\bm A$ is randomly and uniformly chosen from $\{+1,-1\}$.

\begin{figure}[t]
  \centering
  \includegraphics[width=8cm]{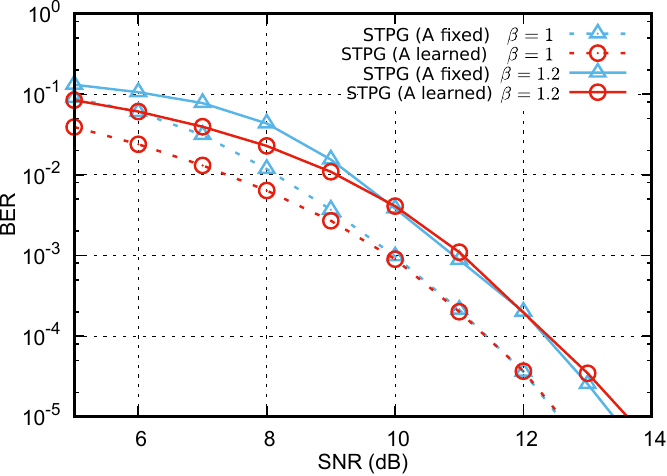}
  \caption{Multiuser detection performance of STPG with/without learning a signature matrix when $\beta=1$ ($m=1200$; dotted lines) and $1.2$ ($m=1000$; solid lines); $n=1200$ and $k=6$.}
  \label{fig_4}
\end{figure}

{Figure~\ref{fig_4} shows BER performance of the STPG detectors with/without learning a signature matrix with overloaded factor $\beta=1$ and $1.2$ when $n=1200$.
It is found that tuning $\bm A$ largely improves detection performance in the low SNR regime in both cases.
When $\beta=1$ and BER$=1.0\times 10^{-2}$, 
the gain of learning a signature matrix is about $0.9$dB. }
On the other hand, the gain vanishes as the SNR increases.
Especially in the case of $\beta=1.2$, 
the joint learning shows worse detection performance than the STPG detector with a fixed signature matrix in the high SNR regime.
This is because the gain of signature design is expected to be small and
training the detector is sensitive to perturbations of a signature matrix when noise level is relatively small.
It is a future task to improve the joint learning method in the high SNR regime.      
These results suggest that the proposed signature design with the STPG detector improves the multiuser detection performance with reasonable training costs especially in the low SNR regime.


\section{Concluding remarks}\label{sec_5}
In this paper, we propose a trainable SCDMA multiuser detector called STPG detector.
Applying the notion of deep unfolding to a computationally efficient PG detector, 
the STPG detector contains a constant number of trainable parameters which can be trained by
standard deep learning techniques.
An advantage of the STPG detector is the low computational cost that is proportional to the number of active users.
Moreover, compared with a conventional BP detector, the STPG detector has less computational complexity with respect to signature sparsity $k$ and signal constellation size while its detection performance is fairly close to that of a BP detector.
In addition, we demonstrate a DL-based signature design using the STPG detector.
Numerical results show that the joint learning method improves multiuser detection performance 
especially in the low SNR regime with reasonable training costs.

\section*{Acknowledgement}
This work was partly supported by JSPS Grant-in-Aid for Scientific Research (B) 
Grant Number 16H02878 (TW) and 
Grant-in-Aid for Young Scientists (Start-up) Grant Number 17H06758 (ST), and the Telecommunications Advancement Foundation (ST).



\end{document}